\documentclass{iucrjournals}

\usepackage{graphicx}
\usepackage{dcolumn}
\usepackage{bm}
\usepackage{mathptmx}
\usepackage{etoolbox}
\usepackage{siunitx}

\title{A high-temperature furnace for multi-modal synchrotron-based X-ray microscopy and diffraction imaging}

\author[a]{Louis Lesage}
\author[a]{Yves Watier}
\author[a]{Helena Isern}
\author[a]{Aditya Shukla}
\author[a]{Virginia Sanna}
\author[a]{Thomas Dufrane}
\author[b]{Yubin Zhang}

\author[a]{Carsten Detlefs}
\author[a]{Can Yıldırım\IUCrCemaillink{can.yildirim@esrf.fr}}
\affil[a]{Experiments Division, European Synchrotron Radiation Facility, 71 Avenue des Martyrs, CS40220, 38043 Grenoble Cedex 9, France.}
\affil[b]{Department of Civil and Mechanical Engineering, Technical University of Denmark, 2800 Kgs. Lyngby, Denmark}

\begin{document} 
\maketitle

\begin{abstract}

The design, calibration, and initial application of a non-contact high-temperature furnace developed for in situ synchrotron X-ray experiments are presented. The system enables a stable operation up to \SI{1000}{\celsius}, with heating rates exceeding \SI{6000}{\celsius\per\minute} and thermal stability better than $\pm \SI{2}{\celsius}$. Temperature calibration was performed using (i) direct measurements with a thermocouple to characterize heating and cooling ramp rates and map temperature gradients along the $x$, $y$, and $z$ axes, and (ii) synchrotron X-ray diffraction to track the ferrite-to-austenite (BCC to FCC) phase transition in an iron grain under beamline conditions.
The furnace's contactless geometry provides full translational and rotational freedom, with \SI{360}{\degree} rotation and wide tilt capabilities, making it fully compatible with a range of diffraction and imaging techniques. Its 3D-printed modular body includes closable apertures for auxiliary functions such as active cooling or X-ray fluorescence. The design is easily customizable for diverse experimental requirements and can be adapted to most beamlines.

The furnace has been implemented at the ID03 beamline of the European Synchrotron Radiation Facility (ESRF) which supports Dark field X-ray Microscopy (DFXM), 3D X-ray Diffraction (3DXRD), magnified topotomography (MTT), phase-contrast tomography (PCT) and diffraction contrast tomography (DCT).
As a first application, a DFXM case study on a cold-rolled Al1050 sample during isothermal annealing is presented. The imaging of a selected grain before and after the heat treatment reveals strain relaxation and grain growth. 
This furnace offers a robust and flexible platform for high-temperature synchrotron studies across materials science, including metals, ceramics, and energy materials. It is now part of the ESRF sample environment pool and is available to all users.

\end{abstract}


\section{Introduction}

The microstructure of a material plays a key role in determining its macroscopic properties and performance in engineering applications. Heat treatments are widely used across industries to tailor these microstructures, thereby tuning mechanical properties such as strength, ductility, and hardness. To better understand how thermal stimuli drive changes in microstructures, in situ and operando studies with sufficient spatial and temporal resolution are essential.

Synchrotron-based X-ray techniques have emerged as powerful, non-destructive tools for investigating such transformations over multiple length scales. Traditional methods, such as scanning electron microscopy (SEM) with electron backscatter diffraction (EBSD) and transmission electron microscopy (TEM), provide excellent spatial resolution but are limited to surface or thin foil analysis and often require invasive sample preparation. In contrast, synchrotron diffraction imaging methods such as three-dimensional X-ray diffraction (3DXRD)~\cite{poulsen2004three}, diffraction contrast tomography (DCT)~\cite{reischig2013advances}, phase-contrast tomography~\cite{Cloetens1999}, and topography~\cite{lang1993early, yildirim2021role} offer non-destructive three-dimensional insight into the internal structure of polycrystalline materials.
Among these methods, Dark-Field X-ray Microscopy (DFXM) has emerged as a powerful technique for high-resolution imaging of orientation and strain within individual grains embedded in bulk crystalline materials, achieved by placing an objective lens in the diffracted beam path~\cite{simons2015dark,yildirim2020probing}.
Following the EBSL2 upgrade~\cite{cloetens2025esrf} of the European Synchrotron Radiation Facility (ESRF), the ID03 beamline~\cite{isern2025esrf} was constructed to host the DFXM technique in conjunction with complementary techniques, including 3DXRD, DCT, diffraction tomography, X-ray topography and phase-contrast tomography, thereby enabling comprehensive multimodal and multiscale investigations of crystalline materials.

Over the past decade, there has been a growing interest in studying the microstructure of crystalline materials under external thermal stimuli using these techniques~\cite{yildirim20224d, ahl2020subgrain, dresselhaus2021situ, yildirim2025pink, mavrikakis2019multi}. 
A previous furnace developed for the ID06 beamline, where DFXM experiments were previously conducted, is no longer compatible with other goniometer setups and faced several technical limitations~\cite{yildirim2020radiation, kutsal2019esrf}. Notably, the furnace had a restricted tilt range and narrow angular apertures for the X-ray beam, both of which are critical for aligning grains to meet diffraction conditions. Although it offered excellent temperature stability, the sample was offset from the goniometer's center of rotation in certain directions, further limiting its suitability for DFXM, where precise angular alignment is essential.
An alternative heating approach involved the use of a gas blower~\cite{yildirim20224d, yildirim2025pink}; however, it lacked the thermal stability required for achieving nanometer-scale resolution. In addition to being costly, it offered limited temperature control, making experiments difficult to reproduce. To address these limitations and support the advancement of synchrotron techniques toward operando and high-throughput studies, a heating solution compatible with the ID03 sample goniometer~\cite{isern2025esrf} is needed, one that combines thermal precision, mechanical flexibility, and experimental reproducibility.

Here, we present a newly developed non-contact furnace that enables full 360-degree sample rotation and a wide tilt range, ensuring seamless integration with advanced goniometer setups. Its 3D printed body enables customization for adaptation to various beamlines and characterization techniques. We report on the furnace's thermal performance, including its temperature homogeneity and ramp rates, based on in situ thermocouple measurements. In addition, we demonstrate how the absolute temperature at the sample position can be determined by monitoring phase transformations and lattice parameter evolution in reference samples. Finally, we showcase the capabilities of the system for in situ studies by capturing grain growth in an aluminum alloy during annealing using DFXM.

\section{\label{sec:spec} Description of the furnace}

\subsection{Conception of the furnace}

\begin{figure*}[ht]
    \centering
    \includegraphics[width=1\textwidth]{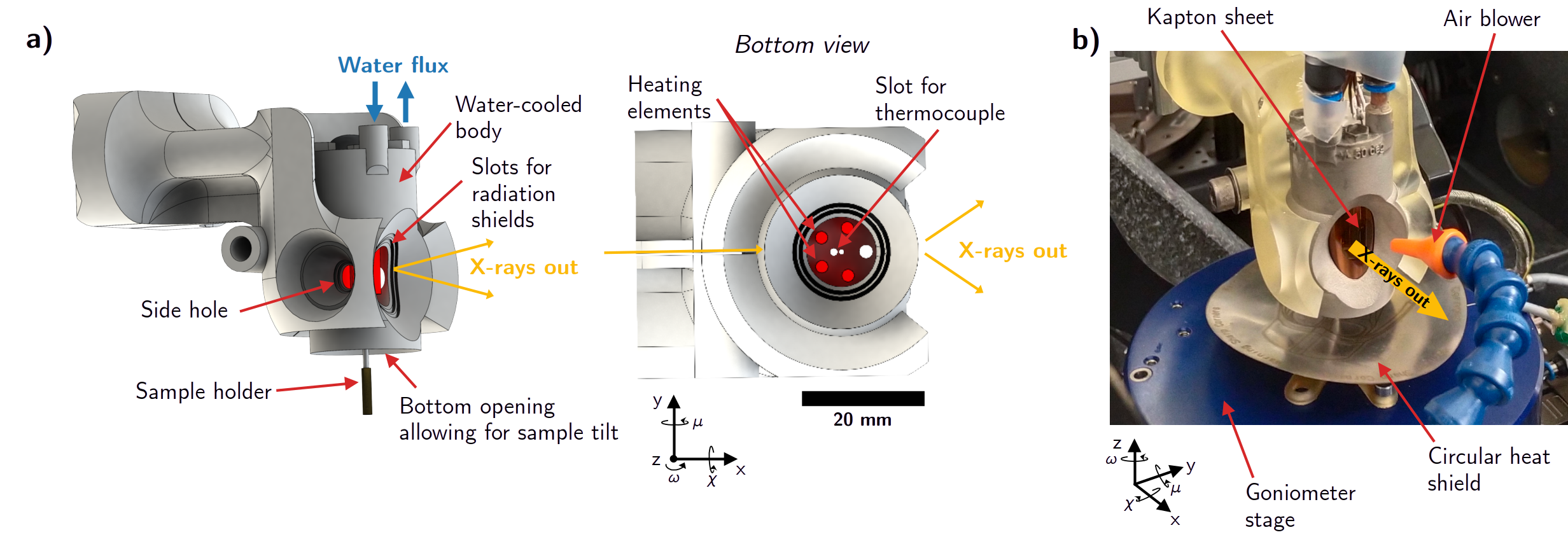}
    \caption{Design and integration of the in situ furnace for synchrotron experiments. (a) Computer-aided design (CAD) rendering showing a side and bottom view of the furnace, with the laboratory coordinate system overlaid in black: x is along the incident X-ray beam direction, z is vertical. Key components such as the sample holder slots, X-ray beam aperture, and water cooling channels are visible. The bottom view shows the placement of the thermocouple and resistive heaters, between which the sample is inserted. (b) Photograph of the furnace mounted on the beamline goniometer. A circular heat shield is inserted on the bottom part to protect the goniometer. The angles $\mu$, $\chi$ and $\omega$ represent goniometer rotations about the laboratory coordinate axes.
}
    \label{fig:furnace}
\end{figure*}

Fig.~\ref{fig:furnace} shows computer-aided design (CAD) images alongside a picture of the furnace in operation.
The furnace's body, fully displayed on Fig.~\ref{fig:furnace}.a, was 3D printed using direct metal laser sintering (DMLS) and made of stainless steel. Such an additive manufacturing technique enables cost-effective replication and customization of the design to accommodate various beamline requirements.
The furnace body is fully water-cooled to maintain a temperature near ambient conditions. This acts as a safety measure in the event of skin contact with the furnace's exterior and helps ensure spatial positioning accuracy by reducing the risk of thermal drift.
To minimize unwanted turbulence and enhance thermal efficiency, the furnace features several printed fixed screens with maximum openings. These are supplemented by laser-cut stainless steel sheets that slide in, serving as radiation shields to reduce openings and decrease air turbulence at the sample position. This design facilitates easy adaptation for various measurements using the same furnace body and simplifies maintenance, as the laser-cut thin metal sheets can be easily replaced.

The furnace body has a small aperture for the incident X-ray beam and a larger opening on the opposite side to allow both transmitted and diffracted beams to exit. This wider aperture supports diffraction angles up to $\pm 50^\circ$, enabling a broad $2\theta$ range. To maintain X-ray transparency while minimizing air turbulence, the exit opening is sealed with a removable Kapton sheet (see Fig.~\ref{fig:furnace}.b). This sheet is air-cooled to prevent melting.
The furnace also includes two side openings that can be used for X-ray fluorescence measurements, thermal imaging, laser probing, or inert gas inflow, demonstrating the system’s versatility. During DFXM experiments, these side holes can be easily sealed with stainless steel covers that are highly reflective to infrared radiation.

Five slots are available to insert heater elements as shown on the bottom view of Fig.~\ref{fig:furnace}.a. They may be all used or not, depending on the required X-ray geometry. For DFXM measurement, a set of four resistances has been inserted into the furnace from its top, and a K-type thermocouple is used to monitor the internal temperature and adjust the heating power to follow the setpoint using a PID temperature controller.

\subsection{Integration to the ID03 beamline}

Two motorized stages allow for precise translation of the furnace along the \( y \) (horizontal, perpendicular to the beam) and \( z \) (vertical) directions, while manual adjustment along the \( x \)-axis (parallel to the X-ray beam) is achieved using a lead screw and guided rails. A reference marker on the rail facilitates reproducible positioning along \( x \). Once the furnace and sample are aligned in the \( x \)–\( y \) plane, the furnace can be translated vertically (\( z \)) to move it out of the beam path without disturbing the sample’s alignment. The furnace is mounted on a separate granite gantry, independent of the sample goniometer, ensuring mechanical stability during high-temperature operations.

The motion of the furnace and the sample are fully decoupled. This design permits sample rotations about the vertical axis ($\omega$ angle) as well as tilts about the horizontal axes ($\mu$ and $\chi$ angles, see Fig.~\ref{fig:furnace} for the definition of the angles). The current setup allows for rotational flexibility of up to \( \pm 25^\circ \) about \( \mu \) and \( \chi \), and full 360° rotation about $\omega$, which is critical for orienting single grains or domains during diffraction imaging experiments and also for tomography measurements. The available translation space exceeds \SI{2}{\milli\meter}, which is sufficient for spatial scans across the sample surface or depth profiling.

The design also accommodates the near-field imaging camera~\cite{isern2025esrf}, a high-resolution detector system based on a PCO sensor coupled with visible-light optics, yielding an effective pixel size of approximately \SI{0.65}{\micro\meter}. This detector can be positioned as close as \SI{50}{\milli\meter} to the sample, even with the furnace in place, due to the compact geometry and top-mounted cabling. The cabling layout ensures that neither the rotational freedom of the goniometer nor the translational motion of the furnace is obstructed. This near-field detector is essential to the measurements performed at the beamline as it is generally used for sample positioning, grain alignment, phase contrast tomography and high resolution rocking curves of a part of the diffraction ring \cite{ahl2020subgrain, lee2024multiscale}.

Finally, the furnace system is integrated with the beamline control software and can be operated remotely via a dedicated BLISS session, eliminating the need to enter the experimental hutch during alignment or temperature ramping. This facilitates efficient and safe operation, especially during long-duration or high-temperature experiments.

This modular and flexible integration makes the system suitable for a wide range of in situ synchrotron experiments, including thermally activated deformation, phase transitions, and recovery studies using advanced diffraction imaging techniques.

\subsection{Temperature distribution within the furnace}

\begin{figure}[ht]
    \centering
    \includegraphics[width=0.45\textwidth]{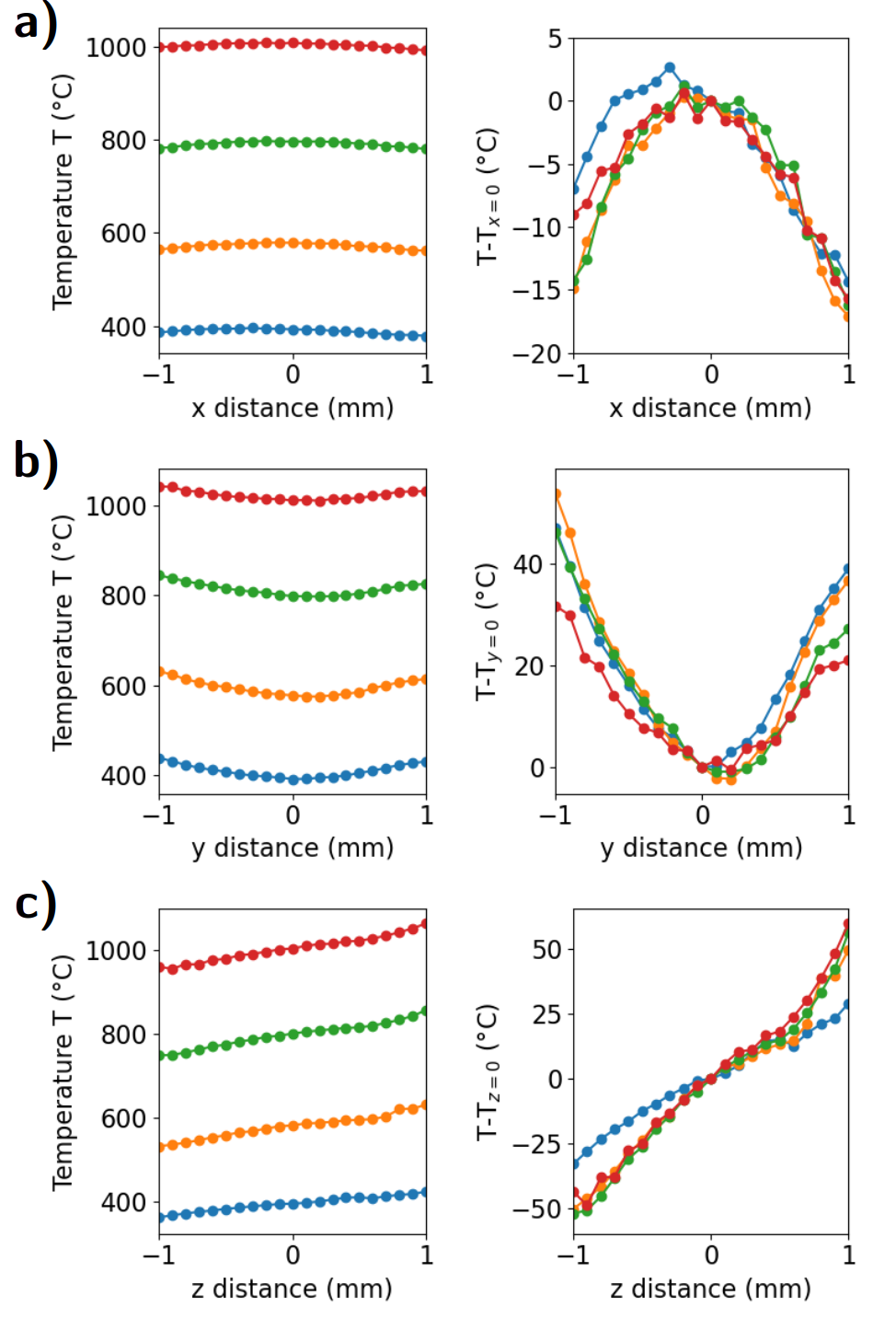}
    \caption{Temperature as a function of the position of the thermocouple-sample in the furnace along (a) $x$-axis, (b) $y$-axis, and (c) $z$-axis. Left panels show the absolute temperature; right panels show the temperature difference relative to the origin position ($x = y = z = 0$). Blue, orange, green, and red curves correspond to set temperatures of 400\,°C, 600\,°C, 800\,°C, and 1000\,°C, respectively.}
    \label{fig:temp_vs_axis}
\end{figure}

To verify the temperature homogeneity inside the furnace, a K-type thermocouple was positioned at the sample position on the goniometer and moved along the laboratory $x$, $y$, and $z$ axes. These temperature measurements are displayed in Fig.~\ref{fig:temp_vs_axis}. Initially, the thermocouple was placed at the origin ($x$=$y$=$z$=0) to align the beam, the inbound and outbound X-ray apertures, and the thermocouple. The initial $x$-position was chosen to be sufficiently distant from the resistances, allowing the sample to be tilted. The $y$ and $z$ origin positions were chosen to align the beam and the tip of the thermocouple.

The results show that for each target temperature (\SI{400}{\celsius}, \SI{600}{\celsius}, \SI{800}{\celsius}, and \SI{1000}{\celsius}), within a region of \SI{\pm 1}{\mm}, the temperature difference relative to the origin position remains below \SI{20}{\celsius} along the $x$-axis, \SI{60}{\celsius} along the $y$ and $z$-axes. This level of variation is expected to have minimal impact on temperature homogeneity, particularly in metallic samples commonly studied with DFXM, due to their high thermal conductivity, which promotes uniform temperature distribution. Additionally, given that the typical field of view in DFXM experiments is approximately \SI{100}{\micro\meter}, it is reasonable to assume that the grain of interest experiences a nearly uniform temperature throughout.

\begin{figure*}[ht]
    \centering
    \includegraphics[width=0.95\textwidth]{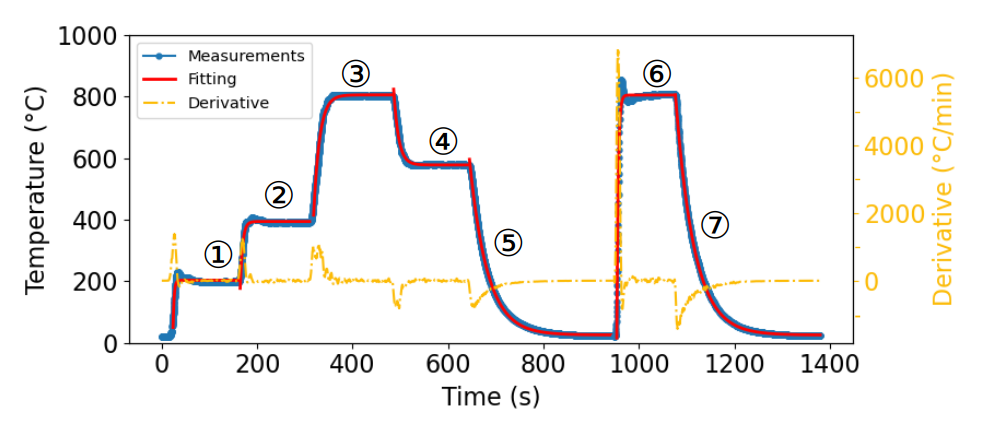}
    \caption{Heating and cooling ramp rates for given temperature setpoints. Red curves refer to fitted curves following Eq.~\ref{eq:exp_decay} for which the fitting parameters are available in Table~\ref{tab:fitting_ramp}. The yellow curve shows the derivative of the fitted heating and cooling curve.}
    \label{fig:temp_ramp}
\end{figure*}

Fig.~\ref{fig:temp_ramp}, steps 1 to 5, shows the heating and cooling rates in a standard operation mode, that is, for a ramp rate set to \SI{1000}{\celsius\per\minute}. It shows that the various targeted temperature plateaus (\SI{200}{\celsius}, \SI{400}{\celsius}, and \SI{800}{\celsius} in steps 1, 2, and 3, respectively) were achieved with the targeted ramp rate. Once the target is reached, the temperature stabilizes with variations of less than $\pm \SI{2}{\celsius}$.
Steps 4 and 5 show a stepwise cooling rate to \SI{600}{\celsius} and further to room temperature. The cooling process relies on natural convection upon turning off the heating resistances or removing the furnace by translating it in the vertical direction. While the furnace was tested here without any active cooling system, its side apertures can accommodate a gas inflow to enhance the cooling rate.

The time evolution of the furnace temperature was fitted using an exponential decay model as in Eq.~\ref{eq:exp_decay}. The fitting parameters are available in Table~\ref{tab:fitting_ramp}.

\begin{equation} \label{eq:exp_decay}
    T(t)=T_0 + A \cdot exp\left({-\frac{t}{\tau}}\right) )
\end{equation}

\begin{table}[ht]
\footnotesize
\caption{Fit parameters for Eq.~\ref{eq:exp_decay} to characterize the temperature ramp using K-type thermocouple. For steps 4 to 7, no ramp rate was set in order to evaluate the maximum heating and cooling rates of the furnace.}
\label{tab:fitting_ramp}
\begin{center}
\begin{tabular}{cccc}
\hline
Temperature Step & Set ramp rate (\SI{ }{\celsius\per\minute}) & $A$ (\SI{ }{\celsius}) & $\tau$ (\SI{ }{\second}) \\
\hline
    1 &  100 & -153 $\pm$ 9 & 2.34 $\pm$ 0.20 \\
    2 &  100 & -217 $\pm$ 5 & 5.11 $\pm$ 0.16 \\
    3 &  100 & -391 $\pm$ 4 & 12.60 $\pm$ 0.23  \\
    4 & Max.  & 246 $\pm$ 2  & 10.34 $\pm$ 0.15 \\
    5 & Max. & 570 $\pm$ 1 & 37.05 $\pm$ 0.11 \\
    6 & Max. & -789 $\pm$ 24  & 2.96 $\pm$ 0.14 \\
    7 & Max.  & 779.55 $\pm$ 2 & 39.04 $\pm$ 0.14 \\
\hline
\end{tabular}
\end{center}
\end{table}

To further characterise the capabilities of the furnace, step 6 consisted of heating the furnace as quickly as possible to \SI{800}{\celsius} without a fixed ramp rate, while step 7 involved the subsequent cooling to room temperature. This operation revealed that the furnace can achieve a heating rate above \SI{6000}{\celsius\per\minute}. However, such a fast ramp rate tends to induce a temperature overshoot, which could be critical for some applications. Adjusting the PID parameters of the furnace may mitigate this overshoot, but would result in a decrease in the ramp rate close to the setpoint.

\section{\label{sec:results} Results}

\subsection{Temperature characterisation from the lattice parameter evolution of iron}

The temperature of the sample, $T_{\text{sample}}$, can differ from the temperature measured by the furnace control thermocouple, $T_{\text{furnace}}$, which regulates the heating setpoint. Since the thermocouple is positioned near the furnace body and not directly on the sample, thermal gradients and response delays may lead to inaccurate estimates of the actual sample temperature. This discrepancy can be critical in temperature-sensitive studies such as phase transformations.

To directly determine $T_{\text{sample}}$, we employed in situ X-ray diffraction to track the evolution of the lattice parameter of iron in its ferritic phase during heating on an area detector.
In addition, iron undergoes a well-known phase transformation from body-centered cubic (BCC) \(\alpha\)-ferrite to face-centered cubic (FCC) \(\gamma\)-austenite at \SI{912}{\celsius} at ambient pressure. This transformation is accompanied by abrupt changes in lattice symmetry and spacing, making it an ideal reference point for temperature calibration. We exploited this transformation as an internal thermometer to calibrate the sample temperature independently of the furnace readout.

The sample, a \SI{1}{\milli\meter} thick square cross-section bar of commercially pure iron, was heated at a rate of \SI{100}{\celsius\per\minute}. Diffraction patterns were recorded every 5 seconds using a FReLoN CCD detector with a resolution of 2048\(\times\)2048 pixels and a pixel size of \SI{47.3}{\micro\meter}, positioned \SI{282}{\milli\meter} downstream of the sample~\cite{isern2025esrf}. The X-ray beam was monochromatic with an energy of \SI{55.12}{\kilo\electronvolt}, selected using a silicon (111) channel-cut monochromator. This high energy allowed simultaneous probing of several diffraction peaks from both the BCC and FCC phases over a broad angular range.

The 2D diffraction images displayed concentric Debye-Scherrer rings, characteristic of polycrystalline materials with randomly oriented grains. These rings were azimuthally integrated over \SI{360}{\degree} using the Python library \texttt{pyFAI}~\cite{kieffer2020new, ashiotis2015fast} to produce 1D diffractograms of intensity versus scattering angle. The detector geometry was calibrated using standard polycrystalline silicon powder. The resulting integrated patterns reveal the evolution of the diffraction peaks during heating.

\begin{figure*}[ht]
    \centering
    \includegraphics[width=0.9\textwidth]{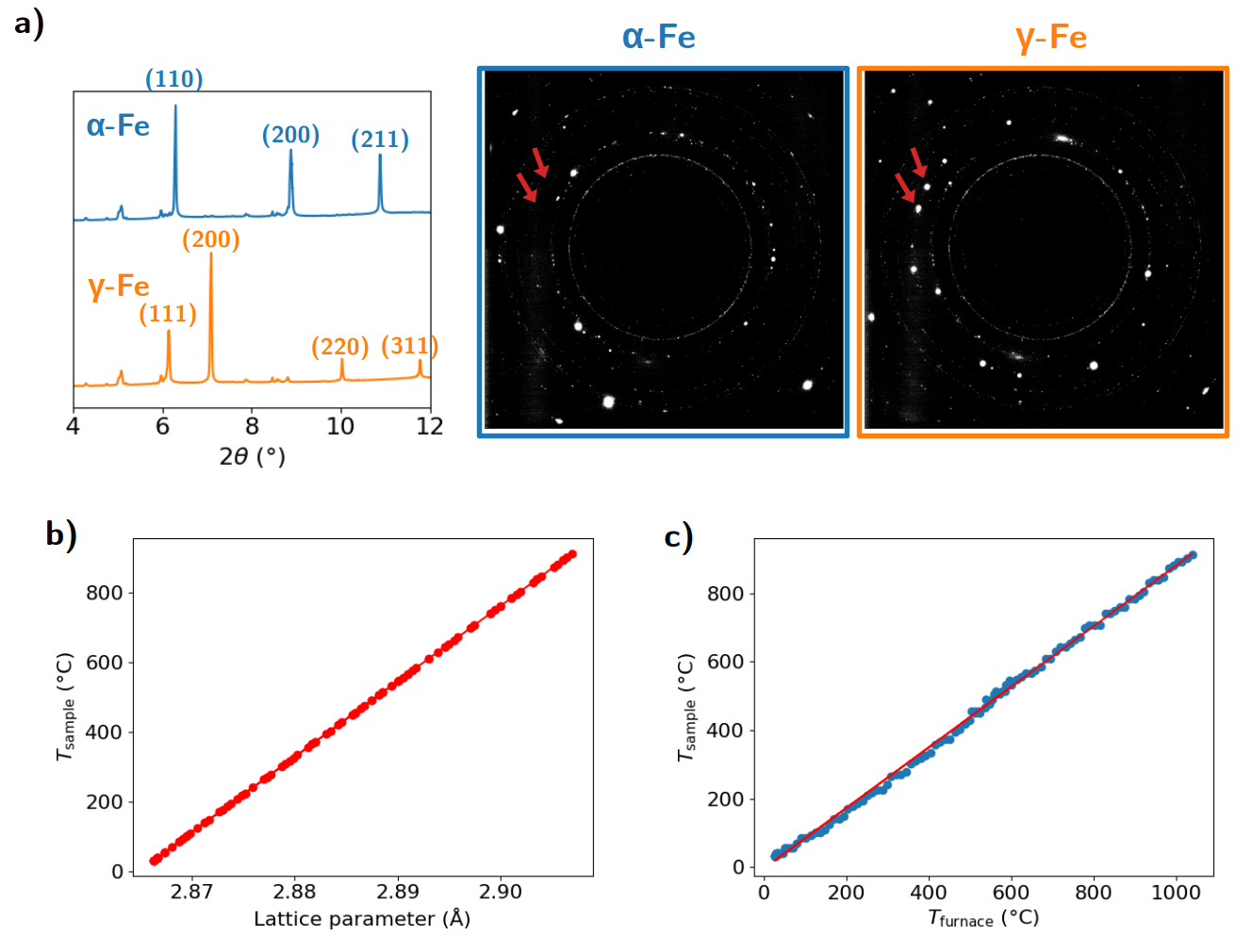}
    \caption{Temperature calibration of the sample through lattice parameter evolution and austeno-ferritic transformation of iron. (a) Integrated 1D diffractograms and detector images for the iron sample before and after the $\alpha$-iron to $\gamma$-iron transition. Red arrows indicate the easily identifiable diffraction spots related to the (200) reflection of austenite, not visible in the ferritic state. (b) Evolution of the sample temperature determined from the lattice parameter evolution of $\alpha$-iron. (c) Sample temperature as a function of the furnace temperature.
    }
    \label{fig:lattice_param}
\end{figure*}

Fig.~\ref{fig:lattice_param}.a shows representative diffractograms before and after the \(\alpha \rightarrow \gamma\) transformation, alongside the corresponding 2D diffraction images. The smaller peaks that are not indexed are attributed to surface oxides. During heating, the BCC peaks shift continuously due to thermal expansion, while new FCC peaks appear abruptly upon transformation. Red arrows mark the emergence of these austenitic reflections, providing a clear signature for identifying the transformation point.

To estimate $T_{\text{sample}}$ during the heating ramp, we extracted the lattice parameter \( a \) of BCC iron from the Bragg peak positions corresponding to the (110), (200), and (211) reflections, using Bragg’s law. Assuming the lattice expands linearly with temperature in the \(\alpha\) phase, and that its value is known at room temperature and at the $\alpha \rightarrow \gamma$ transformation point (\SI{912}{\celsius}), we applied a linear interpolation. Fig.~\ref{fig:lattice_param}.b illustrates this relationship between $T_{\text{sample}}$ and the lattice parameter of $\alpha$-iron. This yields an effective thermal expansion coefficient of $\alpha_L = \SI{16e-6}{\per\celsius}$, consistent with literature values for ferritic iron~\cite{nix1941thermal}.

Fig.~\ref{fig:lattice_param}.c compares the interpolated $T_{\text{sample}}$ with the furnace setpoint $T_{\text{furnace}}$. A clear linear correlation is observed between the two, enabling a practical offset calibration. For a given sample, the exact relationship depends on sample positioning and furnace geometry, this approach should be reliable considering the temperature profile along laboratory translation axes shown in Fig.~\ref{fig:temp_vs_axis}.
To assess stability and reproducibility, we repeated the heating and cooling cycles multiple times. In all cases, the onset of the \(\alpha \rightarrow \gamma\) transformation during heating and the reverse transformation during cooling occurred within \(\pm \SI{5}{\celsius}\) of the previously determined transition temperature. This consistency confirms the reliability of both the furnace system and the diffraction-based temperature calibration. Because the relationship shown in Fig.~\ref{fig:lattice_param}.c depends on the material's absorption properties, and therefore on its nature, it is recommended to perform this calibration prior to any experiment.

In summary, this method offers a robust and material-intrinsic way to determine the true temperature of a sample for in situ synchrotron experiments. By utilizing a well-known phase transformation and tracking lattice parameter evolution with high precision, we provide a reliable internal thermometer. This correction enables more accurate interpretation of structural changes and can be extended to other materials with sharp transformation points.

\subsection{DFXM monitoring of strain relaxation and grain growth in aluminium 1050 alloy}

As a demonstration of the in situ capabilities of the system, we present a study of local orientation changes and grain growth in a 50\% cold-rolled face-centered cubic aluminum 1050 sample during annealing. The specimen, with a cross-section of $500 \times \SI{500}{\micro\meter\squared}$ and a height of \SI{5}{\milli\meter}, was glued to a ceramic stick using heat-resistant adhesive and mounted on the ID03 goniometer. A parallel X-ray beam of size $600 \times 600~\si{\micro\meter\squared}$ was used to illuminate the grain, enabling a 2D projection scan of the local orientation by tilting the sample about the x and y axes ($\chi$ and $\mu$ tilts).
The diffracted X-rays from the (200) reflection were magnified using an objective composed of 87 beryllium compound refractive lenses (CRLs) with a \SI{50}{\micro\meter} radius, positioned \SI{260}{\milli\meter} from the sample. The signal was projected onto a far-field detector located \SI{5346}{\milli\meter} downstream using an indirect detection system comprising a scintillator screen, a 2$\times$ visible light objective, and a PCO.edge camera, yielding an effective pixel size of \SI{175}{\nano\meter}.

The furnace was then positioned around the sample using the near-field camera to ensure that the beam was centered within the inbound X-ray aperture. A grain of the as-received sample was scanned using DFXM prior to the furnace temperature ramp, which was performed stepwise at a rate of \SI{100}{\celsius\per\minute} until visible grain boundary motion was observed in the far-field camera. Based on the $2\theta$ angle of the (200) reflection throughout the ramping, the lattice parameter of the sample, and hence its actual temperature, could be determined using the thermal expansion coefficient of aluminium available in the literature~\cite{wilson1941thermal}. The maximum temperature achieved was \SI{630}{\celsius} and was held for 1 minute. After furnace cooling, the same grain was re-scanned under identical conditions.

\begin{figure}[ht]
    \centering
    \includegraphics[width=0.5\textwidth]{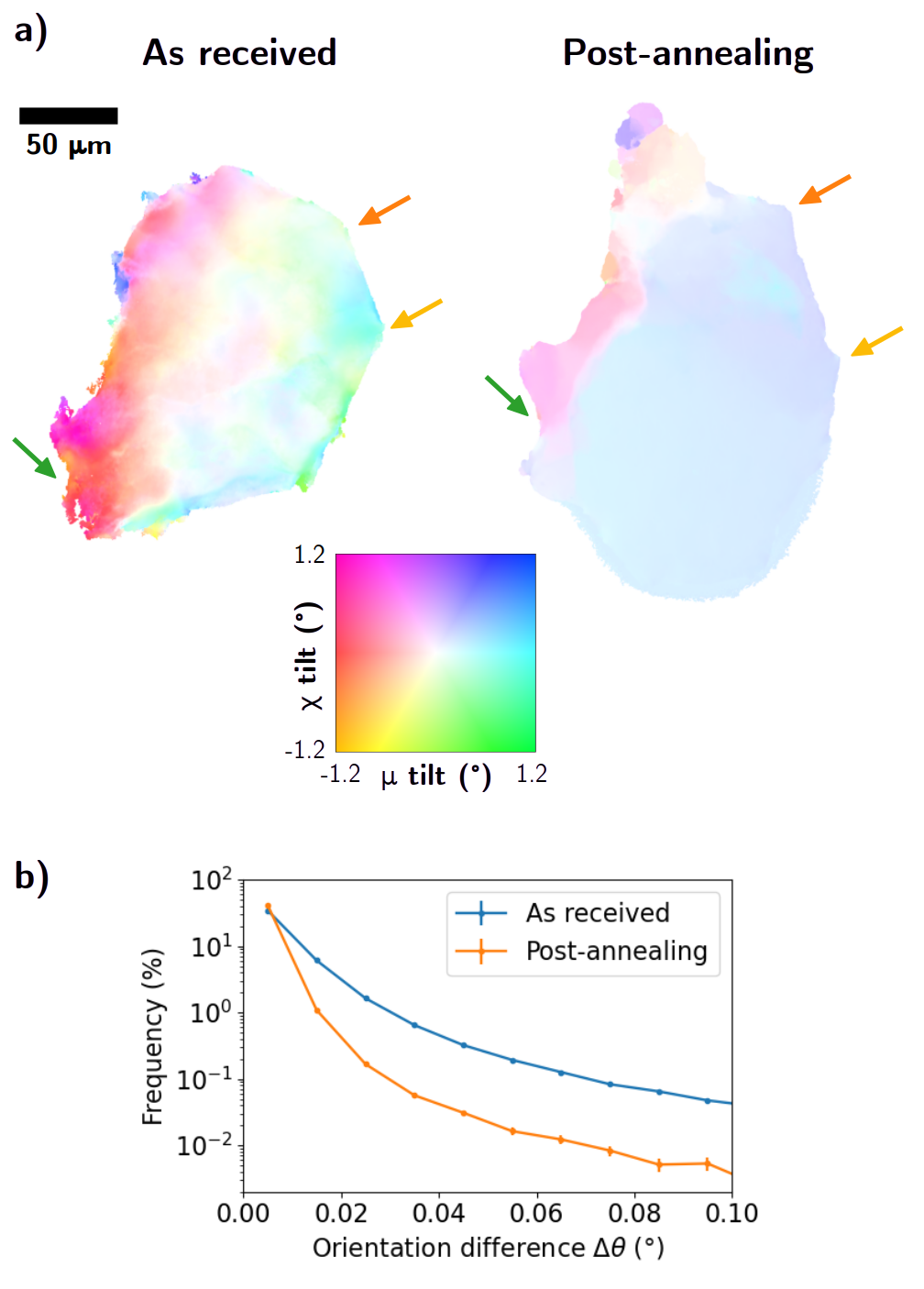}
    \caption{Effect of annealing on the misorientation within a grain. (a) Mosaicity map of a grain from an 50\% cold-rolled Al1050 sample before and after annealing. Arrows indicate distinguishable common features of the grain before and after growth. (b) Frequency of the orientation difference across all pixels of the grain, pre- and post-annealing.}
    \label{fig:mosa}
\end{figure}

Fig.~\ref{fig:mosa}.a presents orientation maps of the sample in its as-received and annealed states, generated from DFXM image stacks after standard preprocessing steps, including background subtraction, region-of-interest selection, and hot pixel removal. Additional details on the data analysis workflow are provided in Ref.~\cite{Garriga2023}. The sample position and tilt range remained nearly identical between the two measurements, ensuring that the same grain was tracked before and after annealing. This is further supported by the identifiable features (indicated by arrows) that persist across both states.

Before annealing, the grain exhibits a clear orientation gradient and heterogeneous internal structure. Following heat treatment, the grain interior becomes significantly more homogeneous, suggesting a reduction in internal mosaicity. This transformation is consistent with a decrease in dislocation density and enhanced recovery processes. 
This observation is quantitatively supported by Fig.~\ref{fig:mosa}.b, which illustrates the distribution of local orientation differences across all pixels of the map. For each pixel, the local orientation difference is calculated as $\Delta\theta = \sqrt{\Delta\mu^2 + \Delta\chi^2}$~\cite{ahl2017ultra}, where $\Delta\mu$ and $\Delta\chi$ represent the average orientation difference with the eight neighboring pixels. Following annealing, the distribution shifts toward lower $\Delta\theta$ values, confirming a significant reduction in intragranular misorientation and a more uniform internal orientation field.

The grain also expanded during annealing. The projected grain area, $A$, calculated by multiplying the number of contributing pixels by the effective pixel area ($175 \times \SI{175}{\nano\meter\squared}$), increased from $A = \SI{21e3}{\micro\meter\squared}$ to $A = \SI{27e3}{\micro\meter\squared}$, corresponding to approximately a 30$\%$ growth. While this measurement captures 2D projected area and does not provide volumetric data, it clearly indicates grain growth driven by grain boundary migration during annealing.
The grain growth observed during annealing appears to be rather directional; the expansion is not isotropic, suggesting that different grain boundaries exhibited varying mobilities. This variation in boundary motion likely reflects differences in boundary character and local driving forces, leading to uneven growth along different interfaces.

Overall, this example demonstrates the capabilities of the furnace setup for \textit{in situ} heat treatment studies using DFXM. The ability to perform annealing without dismounting the sample, while retaining full flexibility in motion and rotation, highlights the system’s effectiveness for tracking microstructural evolution under well-controlled thermal conditions.

\section{Conclusion}

We demonstrated a non-contact radiation furnace developed for versatile X-ray measurement and commissioned for in situ high-temperature X-ray studies at the ID03 beamline of the ESRF. The furnace was calibrated up to \SI{1000}{\celsius} using two independent approaches: direct thermocouple measurements for temperature mapping and ramp rate assessment, and synchrotron diffraction tracking of the ferrite-to-austenite transformation in iron. These methods confirm thermal stability and reproducibility suitable for precise crystallographic experiments. With heating rates exceeding \SI{6000}{\celsius\per\minute} and potential operation above \SI{1000}{\celsius}, the system is well-suited for studies on metallic alloys, ceramics, and functional materials. Though optimized for DFXM, the furnace is fully compatible with additional techniques such as 3DXRD, topotomography, and phase-contrast tomography; it could also be readily adapted to other beamlines. A first application on cold-rolled Al1050 demonstrated the system’s capability to resolve strain evolution during annealing.

Current limitations include the inability to perform rapid quenching or apply mechanical strain during heating. The furnace also lacks a fully sealed atmosphere; while flowing inert gas can mitigate oxidation, strongly oxidizing materials should be enclosed in sealed capillaries. Despite these constraints, the system offers significant flexibility and precision for thermal studies of microstructural evolution across a wide range of synchrotron applications.

\begin{acknowledgements}
We thank ESRF for providing the beamtime at ID03. CY, AS, VS and LL acknowledge the financial support from the ERC Starting Grant n° 10116911.
\end{acknowledgements}

\begin{funding}
This work was supported by the ERC Starting Grant n° 10116911.
\end{funding}

\ConflictsOfInterest{The authors declare that there are no conflicts of interest.
}

\DataAvailability{Raw data were generated at the ESRF large-scale facility. Derived data supporting the findings of this study are available from the corresponding author upon reasonable request.}

\bibliography{furnace}

\end{document}